\documentclass[utf8]{FrontiersinVancouver} % for articles in journals using the Vancouver Reference Style (Numbered), for Frontiers Reference Styles by Journal: https://zendesk.frontiersin.org/hc/en-us/articles/360017860337-Frontiers-Reference-Styles-by-Journal
\usepackage{comment}
\usepackage{url,lineno,microtype,subcaption}

\usepackage[colorlinks=true, linkcolor=black, urlcolor=black, citecolor=black]{hyperref}
\usepackage[onehalfspacing]{setspace}

\def\keyFont{\fontsize{8}{11}\helveticabold }
\def\firstAuthorLast{Kourmpetis {et~al.}} %use et al only if is more than 1 author
\def\Authors{K. Kourmpetis\,$^{1,2}$, P. Laskos-Patkos\,$^{2}$ and Ch.C. Moustakidis\,$^{2*}$}
\def\Address{$^{1}$Theoretical Astrophysics, Eberhard-Karls Universität Tübingen,
D-72076 Tübingen, Germany \\
$^{2}$Department of Theoretical Physics, Aristotle University of Thessaloniki, 54124 Thessaloniki, Greece}
\def\corrAuthor{Ch.C. Moustakidis}

\def\corrEmail{moustaki@auth.gr}

\begin{document}

\onecolumn
\firstpage{1}

\title[CFL matter and light compact stars]{Confronting recent light compact star observations with color-flavor locked quark matter } 

\author[\firstAuthorLast ]{\Authors} %This field will be automatically populated
\address{} %This field will be automatically populated
\correspondance{} %This field will be automatically populated

\extraAuth{}% If there are more than 1 corresponding author, comment this line and uncomment the next one.
%\extraAuth{corresponding Author2 \\ Laboratory X2, Institute X2, Department X2, Organization X2, Street X2, City X2 , State XX2 (only USA, Canada and Australia), Zip Code2, X2 Country X2, email2@uni2.edu}

\maketitle
\begin{abstract}

Recent analyses on the properties of the central compact object in the HESS J1731-347 remnant and the PSR J1231-1411 pulsar indicated that these two compact objects are characterized by similar (low) masses and possibly different radii. This paper aims at reconciling the aforementioned measurements by utilizing the widely employed color-flavor locked (CFL) MIT bag model. The main objective is related to the examination of the acceptable values for the color superconducting gap $\Delta$ and the bag parameter $B$. Furthermore, our analysis involves two distinct hypotheses for the nature of compact stars. Firstly, we considered the case of absolute stability for strange quark matter and we found that it is possible to explain both measurements, while also respecting the latest astronomical constraints on the masses and radii of compact stars. Secondly, we studied the case of hybrid stellar matter (transition from hadrons to quarks), and concluded that, when early phase transitions are considered, the simultaneous reconciliation of both measurements leads to results that are inconsistent to the existence of massive compact stars. However, we showed that all current constraints may be satisfied under the consideration that the HESS J1731-347 remnant contains a {\it slow} stable hybrid star.

\tiny
 \keyFont{ \section{Keywords:} neutron stars, quark stars, color-flavor locked matter, hadron-quark phase transition, hybrid stars, equation of state} %All article types: you may provide up to 8 keywords; at least 5 are mandatory.
\end{abstract}

\section{Introduction}

One of the most important unresolved questions in theoretical astrophysics is related to the nature of matter in the cores of compact stars. Notably, compact stars could be composed solely by hadrons (nucleons and hyperons), but the extreme conditions that prevail in their interior may allow for the presence of exotic forms of matter, such as deconfined quarks~\cite{Witten-1984, Annala-2020}. The latter opens up intriguing scenarios, such as the existence of strange stars~\cite{Weber-2005}, composed purely of quark matter, or hybrid stars~\cite{Heiselberg-2000}, where a quark core is surrounded by a layer of hadrons. Interestingly, given that different hypotheses for the composition of stellar matter may predict distinct properties for the structure of compact stars~\cite{Lattimer-2001,Glendenning-2000}, precisely inferred measurements on masses and radii of compact objects may provide important insight into their nature. 

In 2022, the analysis of Doroshenko {\it et al.}~\cite{Doroshenko-2022} provided puzzling values for the mass ($M$) and radius ($R$) of the central compact object (CCO) in the HESS J1731-347 remnant. More precisely, the authors reported that, at the $1\sigma$ level, $M=0.77^{+0.20}_{-0.17}M_\odot$ and $R=10.4^{+0.86}_{-0.78}$ km. Notably, the surprisingly low mass and radius values, provided by Ref.~\cite{Doroshenko-2022}, led to a wide range of studies attempting their explanation, by considering several different hypotheses for the nature of ultra-dense matter~\cite{DiClemente-2023,Horvath-2023,Oikonomou-2023,Das-2023,Rather-2023,Tsaloukidis-2023,Sagun-2023,Laskos-2024,Laskos-2025,Li-2023n,Mariani-2024,Brodie-2023,Huang-2023,Li-2023,Kubis-2023,Routaray-2023,Char-2024,Tewari-2024}. Specifically, it has been shown that the aforementioned CCO could be a strange quark star~\cite{DiClemente-2023,Horvath-2023,Oikonomou-2023,Das-2023,Rather-2023, Yang-2024, Gholami-2024} (given that there is a mechanism that suppresses rapid cooling, such as color superconductivity~\cite{DiClemente-2023,Horvath-2023}), a hybrid star characterized by an early phase transition~\cite{Tsaloukidis-2023,Sagun-2023,Laskos-2024,Laskos-2025,Li-2023n,Mariani-2024, Gao-2024, Pal-2025}, or a neutron star described by a soft nuclear EOS (at low densities)~\cite{Brodie-2023,Huang-2023,Li-2023,Kubis-2023}. For a relevant critique on the estimation of Ref.~\cite{Doroshenko-2022} and the corresponding required assumptions we refer to the work of Ref.~\cite{AlfordHalpern-2023}.

The recent analysis of Salmi {\it et al.}~\cite{Salmi-2024} also provided intriguing values for the properties of the pulsar PSR J1231-1411. Notably, their results appeared to be sensitive to the selection of the radius prior they used. By limiting the radius to be consistent with previous observational constraints and nuclear theory, the authors indicated that $M=1.04^{+0.05}_{-0.03}M_\odot$ and $R=12.6^{+0.3}_{-0.3}$ km. In addition, by employing an uninformative prior, and constraining the radius between 10 and 14 km, they reported a similar value for $M$ and a higher value for $R$. Interestingly, the recently published work of Qi {\it et al.}~\cite{Qi-2025} provided an alternative estimate on the properties of PSR J1231-1411. In particular, by implementing an algorithm based on the spherical star Schwarzschild spacetime and the Doppler approximation, the authors inferred that the mass-radius properties of the aforementioned pulsar are $M=1.12^{+0.07}_{-0.07}M_\odot$ and $R=9.91^{+0.88}_{-0.86}$ km. In any case, future studies may be required to obtain a more accurate view of this compact object. In the present work, to enrich and extend our investigation, we will consider both estimates of the properties of PSR J1231-1411, referring to the results of Salmi {\it et al.}~\cite{Salmi-2024} as Case I, and to the results of Qi {\it et al.}~\cite{Qi-2025} as Case II.

Given that color-flavor locked (CFL) quark matter \cite{Alford-1999,Alford-2002,Alford-2008} has been successfully employed in the reconciliation of HESS J1731-347, not only for its mass and radius but also for its thermal evolution~\cite{DiClemente-2023,Horvath-2023}, we aim to consider it in order to examine the possible simultaneous explanation of the PSR J1231-1411 properties. To do so, we consider two distinct hypotheses for strange quark matter (SQM): a) absolutely stable, b) energetically favored at high baryon density. Of utmost importance is to examine if the resulting EOSs satisfy other precisely inferred mass-radius measurements (NICER mission)~\cite{Choudhury-2024,Salmi-2024b,Vinciguerra-2024,Riley-2019,Miller-2019} and also allow for the existence of stable massive stars beyond $2M_\odot$~\cite{Antoniadis-2013,Cromatie-2020,Romani-2022}. In this way, we aim to provide a general view on the possible existence of CFL quark matter in compact stars by utilizing all available and accurate observational constraints from rotation-powered millisecond pulsars. The main motivation for performing the present work is analyzed in more depth in the two following paragraphs.

When considering Case I~\cite{Salmi-2024}, it is rather interesting that, while both the CCO in the HESS J1731-347 remant and the PSR J1231-1411 pulsar have similar masses, their radii do not overlap at the $1\sigma$ level. In addition, the fact that PSR J1231-1411 has a slightly higher mass and radius indicates that the $M(R)$ curve may exhibit positive slope ($dM/dR>0$). Notably, this is a standard feature when considering strange stars and hence we aim to constrain the phenomenological parameters (bag constant and color superconducting gap) of the CFL MIT bag model~\cite{Alford-2001,Lugones-2002,Alford-2005}, assuming the absolute stability of strange matter, in light of the aforementioned measurements~\cite{Doroshenko-2022,Salmi-2024,Choudhury-2024,Salmi-2024b,Vinciguerra-2024,Riley-2019,Miller-2019,Antoniadis-2013,Cromatie-2020,Romani-2022}. In any case, this characteristic property ($dM/dR>0$) may not be essential if one considers the possible existence of hybrid stars. In principle, hybrid EOSs, characterized by a large density discontinuity, enable the existence of stars with comparable masses and significantly different radii~\cite{Benic-2015,Alford-2017,Castillo-2019,Blaschke-2020,Christian-2021,Christian-2022,Li-2020,Li-2023c,Li-2023b,Naseri-2024,Jimenez-2024,Li-2025,Zhang-2025}. Therefore, a detailed analysis on the reconciliation of both HESS J1731-347 and PSR J1231-1411 may provide important insight on the existence of low mass twin star solutions (within the CFL MIT bag model).

In Case II~\cite{Qi-2025}, the radius of PSR J1231-1411 has an upper bound (at $1\sigma$) that is even lower compared to the one of the CCO in HESS J1731-347. Taking into consideration that the former pulsar is also more massive than the latter, one can deduce that the simultaneous explanation of both measurements may require a sufficiently soft EOS. In this scenario, it is particularly interesting to examine if the CFL model can also fulfill the conservative $2M_\odot$ maximum mass constraint or support the potential existence of even more massive compact stars like PSR J0952-0607~\cite{Romani-2022}.

This paper is structured as follows: Section \ref{model} sets the theoretical framework of the present study. Section \ref{ReDis} presents the results for the two distinct hypotheses on the nature of compact stars (quark or hybrid stars) along with a detailed discussion on our findings. Lastly, Section \ref{concl} highlights the main insights and conclusions derived from this research.

\section{Color-flavor locked equation of state}
\label{model}

The equation of state for CFL quark matter can be formulated within the MIT bag model framework. To the order of $\Delta^2$ and $m_s^2$, where $m_s$ represents the mass of the strange quark and $\mu$ the quark chemical potential, the pressure and energy density can be expressed as follows $(\hbar=c=1)$ \cite{Lugones-2002,Alford-2005}:
\begin{align}
    P&= \frac{3 \mu^4}{4 \pi^2}a_4+\frac{9 \alpha \mu^2}{2\pi^2}-B \label{eq:eq1}\\
    \varepsilon &= \frac{9 \mu^4}{4 \pi^2}a_4+\frac{9 \alpha \mu^2}{2\pi^2}+B \label{eq:eq2},
\end{align}
where $a_4$ is a parameter that mimics the impact of perturbative QCD (pQCD) corrections \cite{Alford-2005}, $B$ represents the bag constant and $\alpha = -\frac{m_s^2}{6}+\frac{2\Delta^2}{3}$.
% \begin{equation}
%     \alpha = -\frac{m_s^2}{6}+\frac{2\Delta^2}{3}
% \end{equation}
Notably, an analytical expression for $\varepsilon(P)$ can be obtained by combining Eqs. (\ref{eq:eq1}) and (\ref{eq:eq2}):

\begin{equation}
    \varepsilon = 3P+4B-\frac{9\alpha \mu^2}{\pi^2},  \text{ where } \mu^2 =\frac{-9\alpha+ \left[12\pi^2a_4(B+P)+81\alpha^2 \right]^{1/2}}{3a_4} \label{eq:EOS_CFL_h(P)}
\end{equation}
Eq. (\ref{eq:EOS_CFL_h(P)}) reduces to the one presented in Ref. \cite{Lugones-2002} for $a_4=1$, which indicates no strong interactions.

For CFL quark matter to be absolutely stable, its energy per baryon must be lower than the one in the most stable nucleus, $^{56}\mathrm{Fe}$ $(m_b(Fe)=930 \ \mathrm{MeV})$ at zero pressure and temperature \cite{Farhi-1984}. This condition establishes an upper limit for the bag constant as a function of $m_s$ and $\Delta$, for a fixed value of $a_4$:
\begin{equation}
    B<-\frac{m_s^2 m_b^2}{12 \pi^2}+ \frac{\Delta^2 m_b^2}{3 \pi^2} + \frac{m_b^4}{108 \pi^2} a_4\label{eq:CFL_stab_window}
\end{equation}
Conversely, a lower limit can be determined by requiring that two-flavor quark matter should be less stable than nuclear matter. In the context of the MIT bag model $(a_4=1)$, this condition is expressed as \cite{Farhi-1984}:
\begin{equation}
    B \geq 57 \ {\rm MeV \cdot fm}^{-3} \label{B_min}
\end{equation}  
This value decreases when considering lower values for $a_4$. In this study, we use the constraint derived with $a_4=1$ (Eq. (\ref{B_min})), ensuring that all the CFL EOSs used characterize absolutely stable quark matter.

Finally, it is worth noting that in order to ensure that the CFL phase represents the favorable state of matter (compared to 2SC, unpaired or gapless CFL matter) the following condition needs to be met~\cite{Alford-2005b}: 
\begin{equation}
 \Delta \geq m_s^2/2\mu. \label{eq:eq6}   
\end{equation}
Notably, Eq. (\ref{eq:eq6}) holds for all of the quark EOSs constructed in the present study.

\section{Results and Discussion}
\label{ReDis}
\subsection{Strange stars}

In this section, we present our results derived under the assumption that SQM represents the true ground state of matter. However, a key question, that requires a proper discussion, immediately arises: if SQM is more stable than nuclear matter why does normal matter persist?  One explanation may be that nuclear matter is metastable (see Refs.~\cite{Horvath-1992,Oelsen-1992,Iida-1998,Bombaci-2008,Bombaci-2009,Ren-2020} and references therein), separated from the favorable SQM state by a significant energy barrier. Thus, at low densities, the conversion to SQM may be suppressed due to quantum tunneling limitations. However, the extreme neutron star environment, characterized by high densities, may enhance the possibility of a SQM droplet appearing via quantum fluctuations. In addition, extreme events such as supernovae and neutron star mergers may also facilitate SQM formation. Given that the appearance of a SQM seed could trigger the conversion of the entire hadronic star into a strange star, it has long been hypothesized that all compact stars may be of SQM nature.

\begin{figure}[ht]
\centering
\includegraphics[width=\linewidth]{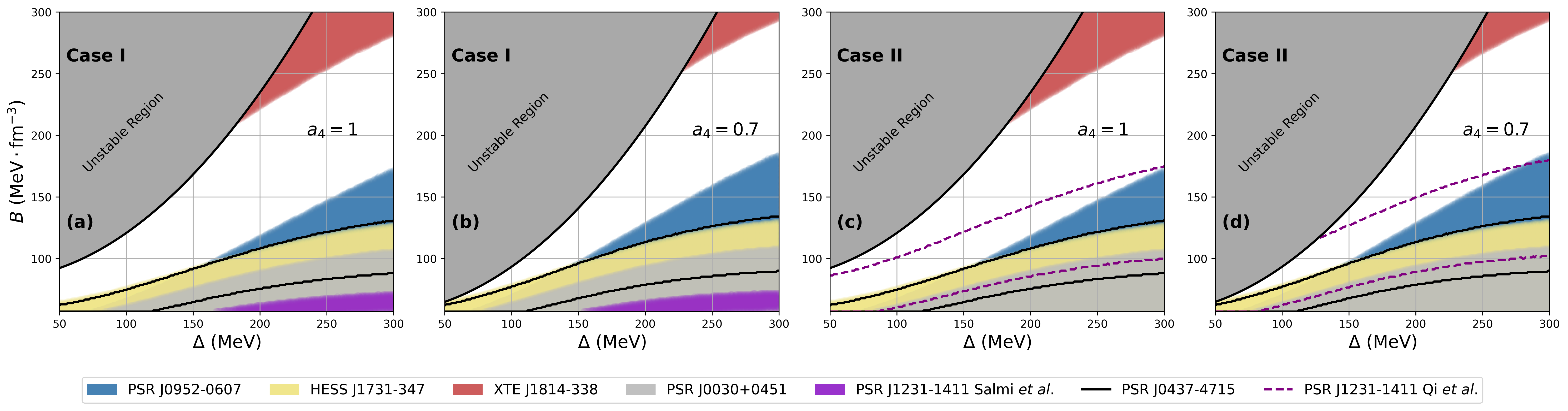}
\caption{$\mathrm{B-\Delta}$ pairs that generate $M-R$ curves that are compatible (at the $1\sigma$ level) with the neutron star observations PSR J0952-0607 \cite{Romani-2022} (blue region), HESS J1731-347 \cite{Doroshenko-2022} (yellow region), XTE J1814-338 \cite{Kini-2024} (red region), PSR J1231-1411 \cite{Salmi-2024} Case I (purple region), PSR J0030+0451 \cite{Miller-2019} (gray region), PSR J0437-4715 \cite{Choudhury-2024} (area between black solid lines) and PSR J1231-1411 \cite{Qi-2025} Case II (area between purple dashed lines). \textbf{(a)} illustrates the parameter spaces when considering Case I and for $a_4=1$, while \textbf{(b)} is identical for $a_4=0.7$. \textbf{(c,d)} examine Case II for $a_4=1$ and $a_4=0.7$ respectively. The dark gray area represents the region where CFL quark matter is not absolutely
stable.}
\label{fig:fig1}
\end{figure}

Proceeding with our calculations, we solved the system of the TOV equations \cite{TOV}, using a wide range of $B \ (57-300 \ \mathrm{MeV \cdot fm^{-3}}$) and $\Delta\ (50-300 \ \mathrm{MeV}$) values to identify the $M-R$ curves that are compatible with the light neutron star measurements for HESS J1731-347 and PSR J1231-1411 (at the $1\sigma$ level). The latter was examined utilizing the results of both Salmi {\it et al.}~\cite{Salmi-2024} (Case I) and Qi {\it et al.}~\cite{Qi-2025} (Case II). Of course, we only considered $B$ and $\Delta$ pairs that lie within the stability window defined by Eq. (\ref{eq:CFL_stab_window}) (fixing the strange quark mass at 95 MeV), while to investigate the possible effects of the $a_4$ parameter we employed two distinct values, $a_4=1$ and $a_4=0.7$. The latter value was selected as it results to EOSs similar to the quark matter EOS from pQCD (e.g., see Refs. \cite{Alford-2005, Fraga-2001}).

Notably, in the present study we work under the assumption of a universal EOS (a single EOS that accounts for all observations). Therefore, the derived EOSs should be compatible to all state-of-the-art multimessenger constraints on the mass and radius of compact stars. Thus, apart from HESS J1731-347 and PSR J1231-1411, we also utilized the corresponding data related to PSR J0952-0607~\cite{Romani-2022}, PSR J0030+0451~\cite{Miller-2019} and PSR J0437-4715~\cite{Choudhury-2024}. Lastly, to highlight the peculiar nature of the recent XTE J1814-338 measurement~\cite{Kini-2024}, we have included it in our analysis, although we did not attempt to interpret it simultaneously with the aforementioned constraints (for some recent works on its reconciliation see Refs. \cite{Zhou-XTE, Yang-XTE, Laskos-XTE, Lopes-XTE, Veselsky-XTE}).

\begin{figure}[ht]
\centering
\includegraphics[width=\linewidth]{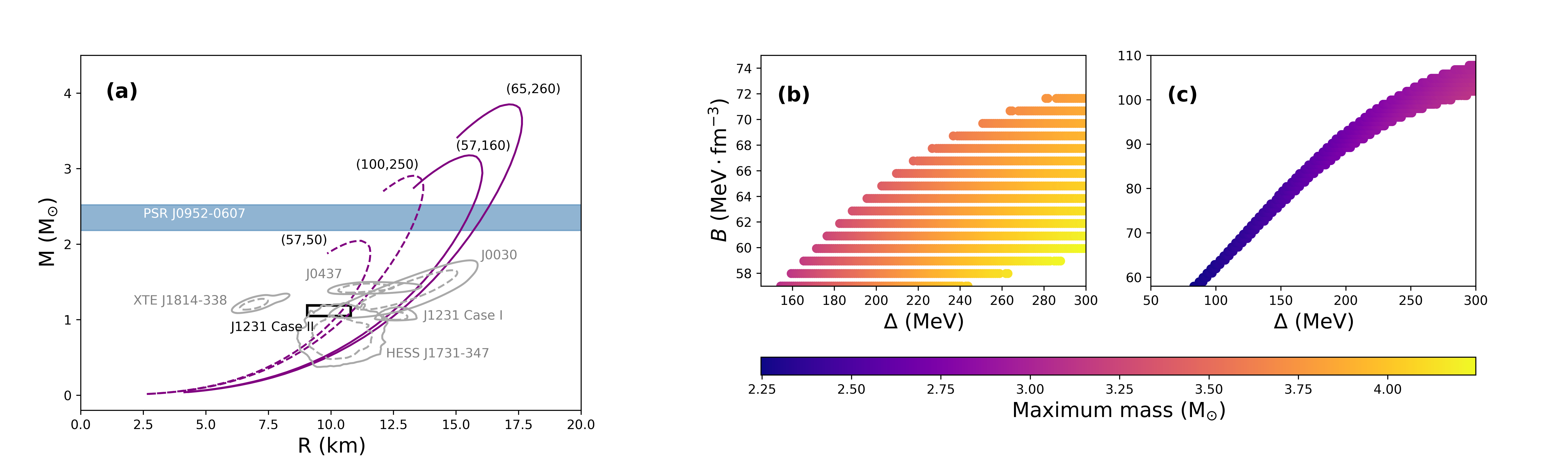}
\caption{\textbf{(a)} Mass-radius curves that are compatible with PSR J1231-1411 Case I (solid lines) and Case II (dashed lines), with $a_4=0.7$. The $B-\Delta$ pairs utilized are representative of the purple region (Case I) and the region between the purple dashed lines (Case II), see Fig. \ref{fig:fig1}(b) and (d). The numbers above each curve represent the $(B,\Delta)$ pairs in $(\mathrm{MeV \cdot fm^{-3}, MeV})$. \textbf{(b)} Close-up of PSR J1231-1411's \cite{Salmi-2024} parameter space considering Case I with $a_4=0.7$, plotted together with the maximum masses reached in this window. Notably, a minimum maximum mass of $\sim3 \ \rm{M_{\odot}}$ is needed. \textbf{(c)} Parameter space that is compatible with PSR J0952-0607 \cite{Romani-2022}, PSR J0030+0451 \cite{Miller-2019}, HESS J1731-347 \cite{Doroshenko-2022}, PSR J0437-4715 \cite{Choudhury-2024} and PSR J1231-1411 \cite{Qi-2025} (Case II) measurements, including the maximum masses reached. This area corresponds to the region between the upper limit of PSR J0030+0451 and the lower one of PSR J1231-1411, see Fig. \ref{fig:fig1}(d).}
\label{fig:fig2}
\end{figure}

Fig. \ref{fig:fig1} displays the $(B,\Delta)$ pairs that are compatible with the above measurements for two different values of $a_4$, accounting for both estimations of PSR J1231-1411's properties. The red, purple, blue, yellow and gray parameter spaces correspond to XTE J1814-338, PSR J1231-1411 (Case I), PSR J0952-0607, HESS J1731-347 and PSR J0030+0451 measurements, respectively. The last three areas extend all the way to the horizontal axis, though it is not visible in the graphs. The region between the black solid lines corresponds to PSR J0437-4715, while the one between the purple dashed lines to PSR J1231-1411 (Case II). Fig. \ref{fig:fig1}(a) illustrates the parameter spaces of the above objects for $a_4=1$ considering Case I, while Fig. \ref{fig:fig1}(b) shows the same graph for $a_4=0.7$. As is evident, although a smaller $a_4$ value alters the CFL EOS, resulting in a modified stability window, it has only minimal impact on the parameter spaces that are compatible to different observations. Finally, in Figs. \ref{fig:fig1}(c) and (d), Case II is examined for $a_4=1$ and $a_4=0.7$ respectively.  

Figs.~\ref{fig:fig1}(a) and (b) indicate that HESS J1731-347 and PSR J1231-1411 (Case I), can both be reconciled within the range of parameters that are compatible to the latter, while also respecting the maximum-mass constraints imposed by PSR J0952-0607.  This region is characterized by small values for the bag constant, close to the minimum possible one, and large values for the color superconducting gap. As a result, the CFL EOS is extremely stiff, leading to very high maximum masses for all $(B,\Delta)$ pairs (above $3\ M_\odot$). The latter is demonstrated in Fig. \ref{fig:fig2}(b), where the maximum masses for all $(B,\Delta)$ combinations of the purple region of Fig. \ref{fig:fig1}(b) are calculated (for the $a_4=0.7$ case).

Another interesting point from Figs.~\ref{fig:fig1}(a) and (b) is related to the fact that the parameter spaces which are compatible with PSR J0437-4715 and PSR J1231-1411 (Case I) do not overlap at the $1\sigma$ level, implying that they cannot be simultaneously explained within this model. However, this tension is mild, as we have checked that a narrow common parameter space exists, when considering the $2\sigma$ estimation for PSR J0437-4715 (see Fig.~\ref{fig:fig2}(a), where we have plotted the $M-R$ diagrams for representative EOSs which are compatible with PSR J1231-1411).

Considering Case II, Figs.~\ref{fig:fig1}(c) and (d) show that there is a small range of parameters which is compatible with all measurements at the $1\sigma$ level (of course except XTE J1814-338). These areas are defined by the lower boundary from PSR J1231-1411 (purple dashed line) and the upper limit from PSR J0030+0451 (gray region). Notably, for PSR J0030+0451 we employed the measurement of Miller {\it et al.}~\cite{Miller-2019}. The aforementioned common areas do not change significantly when adopting the measurements of Riley \textit{et al.} \cite{Riley-2019}, and they would increase when using the lower (regarding the radius) of the three estimations provided by Vinciguerra \textit{et al.} \cite{Vinciguerra-2024}.

The parameter space in which all measurements can be explained by pure CFL matter (Case II) is shown in Fig. \ref{fig:fig2}(c), alongside the corresponding maximum mass predictions (for the $a_4=0.7$ case). The resulting maximum masses span the range of approximately $2.25-3.25 \ \rm{M_{\odot}}$, which is consistent with current observations of heavy pulsars. This maximum mass range, along with that obtained for the parameter space compatible with PSR J1231-1411 (Case I) for $a_4=0.7$, points toward the possible existence of quark stars with masses exceeding $3 \ \rm{M_{\odot}}$. Notably, such masses reside within the so-called \textit{lower mass gap} \cite{Shao-2022,Samsing-2021}, in which the nature of compact objects remains undetermined. Interestingly, evidence for compact objects within the mass gap has emerged from multiple recent studies, based on observations of non-interacting binary systems \cite{Thompson-2019, Jayasinghe-2021}, radio pulsar surveys \cite{Barr-2024}, and gravitational wave detections from binary mergers \cite{Abbot-2020a, Abbot-2023, Abac-2024, Abbot-2017, Abbot-2020b}.

A final remark needs to be made with regards to the possible explanation of the thermal evolution for the CCO in the HESS J1731-347 remnant. Notably, the CCO has a rather high temperature for its estimated age which suggests slow cooling~\cite{DiClemente-2023}, analogous to those of purely hadronic stars. Previous works~\cite{DiClemente-2023,Horvath-2023} have qualitatively suggested that superconductivity may suppress the rapid cooling processes that are expected in unpaired SQM and therefore the temperature of the CCO could be explained. Interestingly, according to Ref.~\cite{Horvath-2023} this could occur only under the consideration of vanishingly small gap values $\Delta<1$ MeV. This is related to the fact that for large values of pairing gaps the suppression of standard cooling processes in the stellar core is dramatic. Therefore, there is a potential tension between the necessity for large gaps, dictated by the large maximum mass constraints, and the low gaps suggested by qualitative analyses on the cooling of the CCO in HESS J1731-347. In addition, it remains an open question whether the low central density expected in the low mass CCO is sufficient for CFL matter to appear for vanishingly small pairing gaps (considering the conditions that need to be met for CFL matter to be energetically favorable~\cite{Alford-2001,Alford-2005b}). In any case, a future precise analysis on the cooling of CFL stars incorporating all of the aforementioned issues, including also potential uncertainties on the estimated temperature and age, is essential to safely conclude about whether the CCO in the HESS J1731-347 could be in fact a CFL strange star.
%%%%%%%%%%%%%%%%%%%%%%%%%%%%%%%%%%%%%%%%%%%%%%%%%%%%%%%%%%%%%%%%%%%%%%%%%

%%%%%%%%%%%%%%%%%%%%%%%%%%%%%%%%%%%%%%%%%%%%%%%%%%%%%%%%%%%%%%%%%%%%%%%%%

\subsection{Hybrid stars}

In the previous section, we considered parametrizations that support the absolute stability of CFL quark matter. At this point, we aim to examine the scenario of explaining all current multimessenger constraints by considering a first-order phase transition, from hadronic to CFL quark matter, in the stellar interior. 

Notably, the simultaneous reconciliation of both HESS J1731-347 and PSR J1231-1431 (in Case I) measurements would be rather difficult assuming a purely hadronic EOS. In particular, the hadronic model should be rather soft, at low densities, to support the low radius associated with HESS J1731-347 in the sub-solar mass region, and then it should rapidly stiffen to achieve the higher radius of PSR J1231-1431 at slightly larger masses. Interestingly, the low-density domain of the nuclear EOS can be effectively constrained through parity-violating electron scattering experiments, which aim to measure the neutron skin thickness of different nuclei. In particular, the PREX-II collaboration provided a measurement for the neutron skin thickness of lead, which pointed to stiff EOS behavior at low densities~\cite{Adhikari-2021,Reed-2021}. Then, the subsequent CREX experiment extracted a puzzling value for the neutron skin thickness of calcium~\cite{Adhikari-2022}, supporting softer models. More precisely, the simultaneous explanation of both CREX and PREX-II was not possible with the use of traditional energy density functionals~\cite{Reinhard-2022,Tagami-2022,Miyatsu-2023,Kumar-2023,Burgio-2024}. However, recent attempts have proposed sophisticated modifications to the Lagrangians which are used to describe nuclear matter, and they achieved an explanation of both experimental values~\cite{Reed-2024,Kumar-2024}. Nonetheless, the predicted EOSs are characterized as extremely stiff. Notably, the latter issue was at some extent resolved in Ref.~\cite{Salinas-2024}.

\begin{figure}[ht]
\centering
\includegraphics[width=\linewidth]{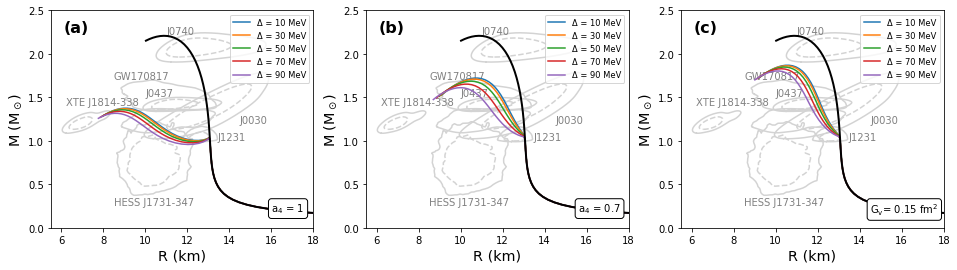}
\caption{Mass-radius diagrams for hybrid EOSs combining the Ska hadronic EOS and the CFL MIT bag model. The gray contour regions denote mass and radius measurements related to PSR J0740+6620 \cite{Salmi-2024b}, PSR J0030+0451 \cite{Miller-2019}, HESS
J1731-347 \cite{Doroshenko-2022}, PSR J0437-4715 \cite{Choudhury-2024} , PSR J1231-1411 \cite{Salmi-2024}, XTE J1814-338 \cite{Kini-2024} and GW170817 \cite{Abbot-2017}. The solid (dashed) contours correspond to $2\sigma \ (1\sigma)$ confidence. \textbf{(a)} shows the results for $a_4=1$. \textbf{(b)} shows the results for $a_4=0.7$. \textbf{(c)} includes the results for $a_4=1$ including the contribution of a vector interaction among quarks with coupling $G_v = 0.15 \ \mathrm{fm^2}$.}
\label{fig:fig3}
\end{figure}

Considering that current experimental constraints might point to stiff behavior for the hadronic EOS at low density, in the present study, we are going to employ a nuclear model which is sufficiently stiff, so that it crosses the PSR J1231-1411 contour (Case I), and, therefore, potentially incompatible to the HESS J1731-347 constraints. In particular, we are going to use a widely employed Skyrme model, namely Ska~\cite{Kohler-1976,Typel-2022,Gulminelli-2015,Danielewicz,Baym-1971}. Thus, in our attempt for explaining all current astronomical measurements, by considering a first-order phase transition to CFL quark matter, we will rely on the hybrid branch to account for the existence of the central compact object in the HESS J1731-347 remnant. It is important to comment that, the employed nuclear model is of nucleonic composition. In principle, as density increases, hyperonic degrees of freedom may appear, altering the properties of the EOS. Given the uncertainties related to hyperon-hyperon and hyperon-nucleon interactions, in the present study we work under a simplified framework that neglects their existence.

In Fig. \ref{fig:fig3}(a), we depicted the mass-radius dependence for hybrid EOSs constructed by varying the values of $\Delta$ and $B$ (fixing $a_4=1$). Notably, the pairs of $\Delta$ and $B$ are selected so that the phase transition density remains fixed, to allow for the hadronic branch to account for PSR J1231-1411. As one can observe, the hybrid branch crosses the HESS J1731-347 inferred region. However, the EOS softening that is induced due to the phase transition plays a critical role on the determination of the resulting maximum mass. As a consequence, the hybrid branch never reaches the $2M_\odot$ and, therefore, it fails to reproduce one of the most robust constraints derived by compact object observations. In Fig. \ref{fig:fig3}(b) we attempted to resolve this issue by lowering the value of the parameter $a_4$. The motivation to do that relies on the knowledge that reducing $a_4$ reduces the value of pressure for a given baryon chemical potential and, as a consequence, for the same parameter values of $B$ and $\Delta$, shifts the onset of the phase transition to higher densities. Therefore, to recover the desired phase transition onset  (i.e., when the hadronic branch crosses the PSR J1231-1411 contour) we need to reduce the value of the bag constant~\cite{Baym-2018}, leading to EOS stiffening. While the constructed EOSs achieve higher masses (still not the conservative $2M_\odot$ limit), the induced stiffening does not allow for an explanation of the HESS J1731-347 measurement. Finally, we tried to stiffen the quark EOS by adding an {\it ad hoc} vector interaction between quarks. We refer to this attempt as {\it ad hoc} since the resulting models do not follow from an initial unified description, but are produced by simply adding a vector mean-field term in the CFL MIT bag model expressed in the canonical ensemble (following Refs.~\cite{Kanakis-2024,Yang-2021,Yang-2023}, the strength of the vector repulsion is quantified via the parameter $G_v=(g/m_v)^2$, where $g$ is the coupling constant and $m_v$ the mass of the mediating boson). For more details on the inclusion of this vector term the reader is referred to Ref.~\cite{Kanakis-2024}. In any case, the results depicted in Fig.~\ref{fig:fig3}(c) indicate that such an approach also does not suffice to reconcile all of the considered astronomical constraints. However, while the CFL MIT bag model might fail in reproducing all measurements (when assuming a phase transition), we can not exclude that a  different, comprehensive analysis (e.g., via a relativistic mean-field type model), that includes such a stiffening term, may enable the simultaneous explanation of all observations.

Up to this moment, we have only considered Case I for PSR J1231-1411. However, it is clear that the consideration of Case II would not alter any conclusion about our inability of reproducing all observational constraints when we combine the CFL model with a stiff hadronic EOS (similar to the employed one).~Nevertheless, future refinement of theoretical models or nuclear experiments that point towards a softer nuclear model may alter the current picture. 

\begin{figure}[ht]
\centering
\includegraphics[width=\textwidth]{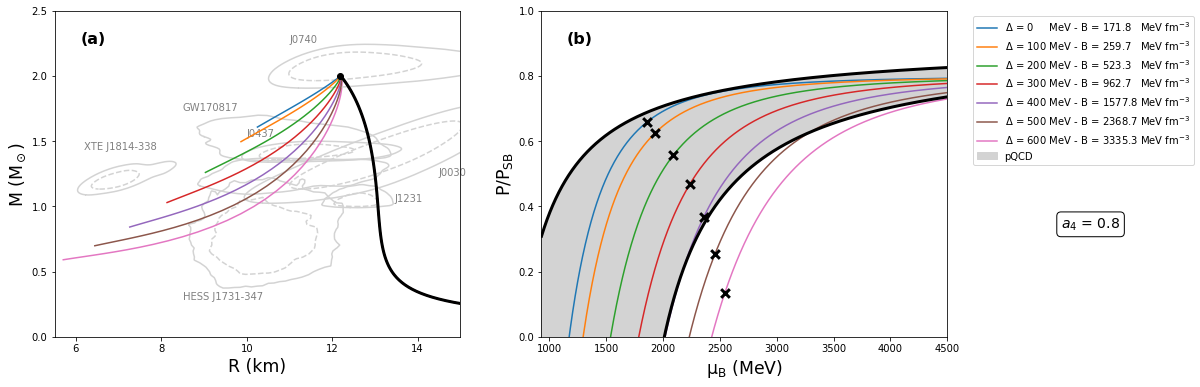}
\caption{\textbf{(a)} Mass-radius diagrams for hybrid EOSs combining the Ska and the CFL bag model with $a_4 = 1$. Parameters $B$ and $\Delta$ are chosen so that the phase transition occurs when the hadronic branch reaches $2 M_\odot$. Each curve is drawn until the last slow stable configuration. The observational data are the same as in Fig. \ref{fig:fig3}. \textbf{(b)} Comparing the non CFL contribution to the quark EOSs to pQCD results \cite{Kurkela-2010,Fraga-2014}. The ‘x’ marks denote the last slow stable configuration.}
\label{fig:fig4}
\end{figure}

Interestingly, we can attempt to reconcile all of the considered astronomical constraints by allowing the hadronic branch to reach the two solar masses and then induce an extremely strong phase transition to make the mass-radius diagram drop to cross the HESS J1731-347 contour. Notably, hybrid stars may be stable even at a descending branch of a mass-radius diagram, assuming that the phase conversion is sufficiently {\it slow}~\cite{Pereira-2018}. The characterization {\it slow} refers to the magnitude of the phase conversion timescale $\tau_\mathrm{conv}$, compared to the period $\tau_\mathrm{osc}$ of radial oscillations that the stellar configurations may undergo. In particular, if $\tau_\mathrm{conv}>>\tau_\mathrm{osc}$, then hybrid stars could be stable even at regions where the well-known turning point criterion would rule them out~\cite{Pereira-2018,Rather-2024}. The idea of explaining the HESS J1731-347 CCO as a {\it slow} stable hybrid star is not new. More precisely, Mariani {\it et al.}~\cite{Mariani-2024} established the plausibility of this scenario in their recent work. In the present study, we aim to examine if the CFL MIT bag model can lead to such an explanation. Specifically, the authors of Ref.~\cite{Mariani-2024} considered a constant speed of sound parametrization where there are three free parameters, namely the transition pressure, the speed of sound and the energy density jump. As a consequence, one can construct a sufficiently stiff EOS, by controlling the speed of sound, which would be characterized by a large energy density jump at any chosen transition density. In contrast, in the current framework we have two free parameters (we are keeping the value of $a_4=0.8$ fixed to account for pQCD constraints; see below) and their interplay has to self-consistently provide an appropriate transition density (so that the hadronic branch reaches the $2M_\odot$), and a large enough energy density discontinuity to trigger an almost vertical drop of the mass-radius curve. Therefore, it is particularly interesting to examine which pairs of the phenomenological parameters $B$ and $\Delta$ could lead to a simultaneous explanation of all current astronomical constraints.

In Fig. \ref{fig:fig4}(a), we depict the mass-radius dependence for hybrid EOSs constructed by varying the color superconducting gap in the range [0,600] MeV. The bag constant was derived by considering that the phase transition occurs when the hadronic mass reaches the $2 M_\odot$. Notably, all of the hybrid branches, in Fig. \ref{fig:fig4}(a), are plotted until the last {\it slow} stable configuration or terminal configuration as named in Ref.~\cite{Mariani-2024}. As one can observe, the reconciliation of the HESS J1731-347 constraints is possible, but it would require extremely large values for both $B$ and $\Delta$ (see the legend of Fig. \ref{fig:fig4}), at least compared to those found usually in the literature. This is, however, somewhat expected as the majority of studies focuses on phase transitions of moderate energy density jumps and low to moderate transition densities. Hence, the extreme conditions required for the explanation of HESS J1731-347 require extreme values for the phenomenological parameters of CFL quark matter. In any case, it is worth noting that recent and robust constraints have been imposed on the color superconduncting gap in Ref.~\cite{Abbott-2025}. While a value of around $\sim$ 400 MeV is potentially compatible to their results, larger values may be disfavored (see Fig. 2 in Ref.~\cite{Abbott-2025}).

Notably, the central baryon chemical potential range that is being considered, in order to achieve the reconciliation of the HESS J1731-347 constraints, reaches such high values that it may cross the regime where the results of pQCD, for the EOS of strongly interacting quark matter, are potentially reliable. Typically, pQCD is considered to be credible at densities around $\sim 40n_0$ (where $n_0$ is the nuclear saturation density), which correspond to values of baryon chemical potential that most likely exceed 2 GeV~\cite{Abbott-2025,Kurkela-2010,Fraga-2014,Kurkela-2014,Annala-2018,Annala-2020,Kurkela-2024}. For that matter, we wanted to test whether the constructed EOSs are compatible to pQCD results (for values beyond 2 GeV), at least for values of baryon chemical potential equal to those found at the center of the last {\it slow} stable configurations. In Fig. \ref{fig:fig4}(b), we plotted the quark EOSs, compared to $\mathcal{O}(a_s^2)$ results of pQCD~\cite{Kurkela-2010} (employing the fit to the numerical data provided by Fraga {\it et al.}~\cite{Fraga-2014}). Notably, each curve depicts the pressure (divided by the Stefan-Boltzmann pressure $P_{SB}=3(\mu_B/3)^4/(4\pi^2\hbar^3)$) as a function of the chemical potential, neglecting the contribution of color superconductivity (as such a contribution is not encapsulated in the considered pQCD calculations~\cite{Fraga-2014}). Finally, the 'x' appearing on each curve denotes the terminal configuration for each EOS. As is evident, for values of $\Delta$ up to $400$ MeV the derived EOSs are compatible to the results of pQCD. However, this is not the case for larger $\Delta$ values. The latter is of utmost importance as it places strong constraints on the possible range of parameters that may enable the explanation of the HESS J1731-347 constraints in a {\it slow} phase conversion scenario.

A final remark is appropriate regarding the fact that while the existence of {\it slow} stable hybrid stars is theoretically intriguing it is not clear how such objects are born. In that sense, future work on possible formation scenarios of such objects would be of utmost importance. Some recent progress on the astrophysical paths that may lead to the existence of twin stars (but for a hybrid branch respecting the turning point criterion) has been made in the work of Ref.~\cite{Naseri-2024}. In that direction, we expect that future research will hopefully shed light on the possible existence of {\it slow} stable hybrid star branches.

\section{Conclusions}
\label{concl}

In this paper, we have presented a systematic study of the CFL MIT bag model in light of recent observations of low-mass compact stars. We have shown that the intriguing measurements of HESS J1731-347 and PSR J1231-1411 can be simultaneously explained within the framework of pure (absolutely stable) CFL matter, while also satisfying the maximum mass constraint set by PSR J0952-0607 and the latest multimessenger constraints on compact star masses and radii (PSR J0030+0451 and PSR J0437-4715). The parameter space consistent with all these measurements yields $M-R$ relations with maximum masses potentially exceeding $3M_\odot$, a result that opens the discussion of quark stars populating the \textit{lower mass gap}.

Notably, in the case of absolutely stable SQM, we worked under the assumption that only one state of matter may appear in the stellar interior. However, it is worth mentioning that a new exotic scenario has been reported in the literature suggesting the possible reappearance of hadrons, at large densities, even if SQM represents the true ground state of matter~\cite{Zhang-2023, Negreiros-2025, Zhang-2024}. Thus, an interesting direction for future work would be to investigate how such a consideration might alter the results reported in the present study.

When considering the framework of hybrid stars, we found that although a hybrid branch originating in the PSR J1231-1411 mass-radius region (as provided by Ref.~\cite{Salmi-2024}) can also accommodate the HESS J1731-347 measurement, the resulting maximum masses remain well below the well-established threshold of $2M_\odot$. While altering the pQCD related parameter $a_4$ or incorporating a repulsive vector interaction increases the predicted maximum mass, the induced stiffening does not allow for the explanation of the HESS J1731-347 constraints (assuming a stiff hadronic EOS).
However, the situation changes if the CCO in the HESS J1731-347 remnant is interpreted as a \textit{slow} stable hybrid star. In this scenario, we adopted a sufficiently strong phase transition at high central pressures, requiring high values for both $\Delta$ and $B$, to produce a hybrid branch that initiates at $2M_\odot$ and drops thereafter, while maintaining stability. We concluded that such a hybrid EOS can be compatible with all of the aforementioned measurements, while also satisfying the demands of pQCD at extremely high density.

\section*{Acknowledgments}
The authors would like to thank
Mr Y. Kini for providing the data for the contour regions
for XTE J1814-338.
P. L.-P. acknowledges that the research work was supported by the Hellenic Foundation for Research and Innovation (HFRI) under the 5th Call for HFRI PhD Fellowships (Fellowship Number: 19175).

\small
% \footnotesize

\end{document}